\documentclass[a4paper,twoside]{article}

\usepackage{epsfig}
\usepackage{subcaption}
\usepackage{calc}
\usepackage{amssymb}
\usepackage{amstext}
\usepackage{amsmath}
\usepackage{amsthm}
\usepackage{multicol}
\usepackage{pslatex}
\usepackage{apalike}
\usepackage{multirow}
\usepackage[dvipsnames]{xcolor}

\usepackage{SCITEPRESS}     
\usepackage[a4paper, left=2.6cm,right=2.6cm,top=3.3cm,bottom=4.2cm] {geometry}
\begin{document}

\title{Advancing Eosinophilic Esophagitis Diagnosis and Phenotype Assessment with Deep Learning Computer Vision}

\author{\authorname{William Adorno III\sup{1}, Alexis Catalano\sup{2,3}, Lubaina Ehsan \sup{3}, Hans Vitzhum von Eckstaedt\sup{3}, Barrett Barnes\sup{4}, Emily McGowan\sup{5}, Sana Syed\sup{4*}, and Donald E. Brown\sup{6*}}
\affiliation{\sup{1}Dept. of Engineering Systems and Environment, University of Virginia, Charlottesville, VA, USA}
\affiliation{\sup{2}College of Dental Medicine, Columbia University, New York City, NY, USA}
\affiliation{\sup{3}School of Medicine, University of Virginia, Charlottesville, VA, USA}
\affiliation{\sup{4}Department of Pediatrics, School of Medicine, University of Virginia, Charlottesville, VA, USA}
\affiliation{\sup{5}Department of Medicine, University of Virginia, Charlottesville, VA, USA}
\affiliation{\sup{6}School of Data Science, University of Virginia, Charlottesville, VA, USA}
\affiliation{\sup{*}Co-corresponding Authors}
\email{\{ss8xj, deb\}@virginia.edu}}


\abstract{Eosinophilic Esophagitis (EoE) is an inflammatory esophageal disease which is increasing in prevalence. The diagnostic gold-standard involves manual review of a patient's biopsy tissue sample by a clinical pathologist for the presence of 15 or greater eosinophils within a single high-power field (400$\times$ magnification). Diagnosing EoE can be a cumbersome process with added difficulty for assessing the severity and progression of disease. We propose an automated approach for quantifying eosinophils using deep image segmentation. A U-Net model and post-processing system are applied to generate eosinophil-based statistics that can diagnose EoE as well as describe disease severity and progression. These statistics are captured in biopsies at the initial EoE diagnosis and are then compared with patient metadata: clinical and treatment phenotypes. The goal is to find linkages that could potentially guide treatment plans for new patients at their initial disease diagnosis. A deep image classification model is further applied to discover features other than eosinophils that can be used to diagnose EoE. This is the first study to utilize a deep learning computer vision approach for EoE diagnosis and to provide an automated process for tracking disease severity and progression.}

\onecolumn \maketitle \normalsize \setcounter{footnote}{0} \vfill

\section{\uppercase{Introduction}}
\label{sec:introduction}

\noindent Eosinophilic Esophagitis (EoE) is a chronic allergen/immune disease that occurs when eosinophils, a type white blood cell, concentrate in the esophagus. It occurs in roughly 0.5 - 1.0 in 1,000 people and is found in 2 - 7\% of patients who undergo endoscopy for any reason \cite{dellon2014epidemiology}. EoE is believed to be triggered by dietary components in patients and is increasing in prevalence \cite{carr2018eosinophilic}. Clinical symptoms include swallowing difficulties, food impaction, and chest pain \cite{runge2017causes}. Persistent esophageal inflammation can eventually progress to strictures in the esophagus with the need for interventional procedures such as dilations, which can significantly impact the quality of patients' lives. The most characteristic microscopic pathologic feature used to diagnose EoE is intraepithelial eosinophil inflammation. For diagnosis, patients with clinical symptoms concerning for EoE undergo an endoscopy and the collected biopsy tissue samples are then evaluated for presence of eosinophils. The accepted criterion for pathologists to diagnose EoE involves identifying at least one High-Power Field (HPF; $400\times$ magnification adjustment) within a patient’s tissue biopsy slide that contains 15 or more eosinophils  \cite{furuta2007eosinophilic}. 

Due to the workload required, pathologists typically only collect information required for diagnosis rather than extensively counting and characterizing eosinophil presence for entire biopsy Whole-Slide Images (WSI). Furthermore, gastroenterologists are currently unable to predict a patient's risk of EoE progression, their clinical phenotype, or the most appropriate treatment plan using available baseline biopsies. Manual quantitative measurements of mean counts of eosinophils in several or all HPF in esophageal biopsies are currently used for research purposes. \cite{dellon2014phenotypic,godwin2020eoe}. Due to the difficulty of manual evaluation of slides, computer automation of detecting and counting eosinophils can efficiently assist pathologists for not only diagnosing EoE and determining severity, but also pave the way for clinically meaningful linkages with EoE clinical and treatment phenotypes \cite{reed2018prolonged}. 

In this paper, we present a complete approach to create and validate an automated eosinophil detection model. Development of the model required annotated data representing the true locations of eosinophils within biopsy image patches. Medically-trained technicians manually annotated images to support the approach due to a lack of publicly available Hematoxylin and Eosin (H\&E) stained biopsy images with annotated eosinophils. With these annotations, a deep learning Convolutional Neural Network (CNN) model was trained to predict the location of eosinophils on new images. The U-Net architecture, which has been originally designed for biomedical image segmentation, is well-suited for eosinophil detection \cite{ronneberger2015u}. After the segmentation model was developed, the prediction masks were post-processed to develop and extract eosinophil counts and statistics. These statistics were then used to quantify and describe EoE severity and its phenotype in patients.

These generated statistics can also reveal linkages between features within initial biopsies and how they correspond with the eventual clinical phenotype and optimal treatment plan. Currently, neither a patient's subsequent clinical phenotype nor the most effective treatment plan can be assessed at the time of initial biopsy and diagnosis. Due to this knowledge gap, patients often undergo algorithmic trials of various treatments before the most suitable treatment is found. The eventual goal of this research is to recommend treatment plans with a higher likelihood of effectiveness and to assess the risk of developing a certain EoE clinical phenotype at the time of initial diagnosis. The various treatment plans include: steroid responsiveness, dairy elimination, and 4-to-6 food elimination. EoE severity and extent statistics are also assessed to find initial biopsy linkages with clinical phenotypes: inflammatory vs. stricturing vs. Proton Pump Inhibitor-responsive esophageal eosinophilia (PPI-REE). Even though PPI is a type of treatment, there have been studies describing a difference in underlying immune and antigen-driven responses in patients with PPI-REE vs. the rest due to which have included it as a separate clinical phenotype \cite{liacouras2011eosinophilic,wilson2018diagnosis}. 

While there are some other known pathologic features in EoE biopsies, there is currently only a manual, categorical scoring system available \cite{collins2017cincyeoe}. To explore this we also applied a classification Convolutional Neural Network (CNN) using WSI patches from EoE-diagnosed and histologically normal patients. This analysis revealed that EoE diagnosis can be highly accurate even when examining small areas of tissue that do not contain a large amount of eosinophils. Both applications of deep learning, image segmentation and classification, provide an increased understanding of EoE and how tissue characteristics can impact diagnosis, severity, treatment plans, and clinical phenotypes. These models combined with an influx of patient response data will pave the way for improved patient outcomes.  

\begin{table}[htbp]
\caption{Table of Featured Acronyms}
\begin{center}
\begin{tabular}{|c|c|}
\hline
\textbf{Acronym} & \textbf{Definition} \\
\hline
CNN & Convolutional Neural Network \\
\hline
EoE & Eosinophilic Esophagitis \\
\hline
FED & Food Elimination Diet \\
\hline
GPU & Graphics Processing Unit \\
\hline
Grad- & Gradient-weighted \\
CAM &  Class Activation Mapping \\
\hline
H\&E & Hematoxylin and Eosin \\
\hline
HPF & High-Power Field \\
\hline
PPI & Proton Pump Inhibitor \\
\hline
REE & Responsive Esophageal Eosinophilia \\
\hline
WSI & Whole Slide Image \\
\hline
\end{tabular}
\label{tab-acronyms}
\end{center}
\end{table}

\section{\uppercase{Background}}

\noindent Previous research for automated eosinophil detection focused on the image classification of these cells, among others, in blood specimens \cite{liang2018combining}. In contrast, the presence and appearance of eosinophils in H\&E stained biopsy tissue samples is different from those present in the blood since they are embedded with the tissue and are surrounded by various other cellular structures with overlapping color variations and gradients. Objects are freely floating in blood images where there are stark differences between the white and red blood cells vs. uni-colored background. The strategy for prediction of white cell blood types is to isolate a single cell per patch and perform image classification \cite{habibzadeh2018automatic,ozyurt2019fused}. This approach is difficult for large WSIs since it will require processing thousands of small patches for detecting and counting each eosinophil. White blood cell segmentation has previously also been executed using only 42 cropped images via a SegNet model, which achieved high accuracy \cite{tran2018blood}. 

Cell segmentation has been shown to be a critical component of linking biopsy features with disease diagnosis and outcomes. Nuclei segmentation recently increased in popularity with publicly available datasets such as Kaggle's 2018 Data Science Bowl and the Multi-organ Nuclei Segmentation Challenge \cite{kumar2019multi}. Eosinophils have unique features such as a bi-lobed nuclei and cytoplasm filled with red to pink granules \cite{Rosenberg2012}. The complexity of this segmentation is why a deep learning approach is preferred over simpler techniques. The U-Net CNN architecture was originally designed for biomedical image segmentation to produce high-resolution segmentation masks without requiring a large amount of training data \cite{ronneberger2015u}. U-Nets are fully-convolutional models that contain a series of convolutional blocks to contract and then expand the image to output a segmentation mask of the same dimensions as the input. Improvements to the original U-Net architecture were developed over the past few years including Residual U-Net (Res. U-Net). It incorporates residual blocks into the architecture to increase the depth of the model while still propagating information quickly \cite{zhang2018road}. RU-Net and R2U-Net models also introduced recurrent convolutional layers to improve feature representation \cite{8556686}. Attention gates were further incorporated into U-Nets for highlighting important features that pass through skip connections \cite{oktay2018attention}.

There are various different approaches for the segmentation model loss function with the two major classes being per-pixel or per-image metrics. It has been recommended to train the model using the same metric as the one used for evaluation rather than combining per-pixel metrics with per-image metrics in the loss function \cite{eelbode2020optimization}. Per-image metrics such as region-based techniques tend to perform better with mild class imbalance \cite{sudre2017generalised}. In our case, there was a large imbalance of more background pixels than foreground (annotated eosinophils). Dice coefficient loss has been shown to be a popular region-based approach with successful results for class imbalanced problems \cite{milletari2016v}. 

For image classification, state-of-the-art architectures have been developed through the ImageNet Large Scale Visual Recognition Challenge \cite{russakovsky2015imagenet}. The VGG16 architecture is not the most recent CNN innovation, but it has shown to perform effectively and has large image dimensions in the last convolutional layer compared to most other ImageNet models \cite{Simonyan15}. The dimensions of the last convolutional layer can impact the resolution of the Gradient-weighted Class Activation Mapping (Grad-CAMs). Grad-CAMs are visual explanations of CNNs that highlight the important regions of the image using the gradients that flow into the last convolutional layer \cite{selvaraju2017grad}. Grad-CAMs have a similar appearance to heat maps overlaid on an input image and are used to identify cellular-level features that are important for diagnosis models. Grad-CAMs are typically used to confirm whether the features being utilized by the models for classification are relevant rather than being extraneous parts of images.

\section{\uppercase{Methodology}}

\noindent In this section, we discuss the methods for the image segmentation of eosinophils and image classification for EoE diagnosis.

\subsection{Image Segmentation of Eosinophils}

\subsubsection{Data Generation}

The WSIs were obtained via digitization of archived biopsy slides present at our center for patients who had already been diagnosed with EoE. This was done via scanning of biopsy tissue slides using a Hamamatsu NanoZoomer S360 Digital slide scanner C13220 with the scanner setting being maintained at $40\times$ magnification. A pool of 274 512 $\times$ 512 patches were generated from WSIs of seven different EoE patients. This image size provided a reasonable annotation area for outlining eosinophils, while also providing a practical input size for segmentation models. A total of 1,037 eosinophils were annotated by a biology trainee who closely supervised by a gastroenterologist and a pathologist. Annotations were done via manually creating polygons around eosinophils using Aperio Imagescope. These patches were smaller than the HPF (conversion of HPF to pixels explained below) used for EoE diagnosis, but the model was still able to detect eosinophils within an HPF using a sliding window approach. An HPF is of $400\times$ magnification or $40\times$ objective and $10\times$ ocular lenses. 

During real-time histopathological diagnosis using microscopes, the $400\times$ HPFs appear circular and have an area of 0.21mm$^2$ \cite{nielsen2014optimal}. We calculated the dimensions of a square patch of the same area to simplify eosinophil counting for deep convolutional models. First, the square root was computed to find the length of the sides of square in metric units: $\sqrt{0.21} = 0.46$mm or 460 microns. The scale of the EoE biopsy images was approximately 0.23 microns per pixel. Therefore, the length of one side of a square HPF was $\frac{460}{0.23} = 2,000$ pixels. To generate HPF samples from WSI, 2,000 $\times$ 2,000 pixel patches were then generated.

\subsubsection{Model Selection and Training}

In addition to the traditional U-Net model, we evaluated the performance of the Residual U-Net, R2U-Net, and Attention U-Net. A 6-fold cross validation was used to ensure that accuracy is not biased towards a small test set. Each model could vary in the number of total parameters via adjustment of the number of filters in the initial convolutional block. Then, the number of filters was doubled in each convolutional block on the contracting side until the tip of the ``U'' was reached. We tested architectures with increasing numbers of filters and parameters, but were eventually restricted by GPU memory limits. Each model was trained using the Adam optimizer for 200 epochs with a learning rate of $2\times10^{-5}$ \cite{kingma_adam}. The model weights that achieved the highest validation accuracy were used for testing. Table \ref{tab-crossval} shows the results from the cross validation. The \textit{size} field in Table \ref{tab-crossval} refers to the total number of parameters of each model. 

\begin{table}[htbp]
\caption{Cross Validation for Model Selection}
\begin{center}
\begin{tabular}{|c|c|c|c|c|}
\hline
\multicolumn{5}{|c|}{\textbf{Test Set Dice Coefficient Statistics}} \\
\hline
\textbf{Model} & \textbf{\textit{Size}}& \textbf{\textit{Median}} & \textbf{\textit{Min}} & \textbf{\textit{Max}} \\
\hline
U-Net & 4.9M & 0.628 & 0.594 & 0.685  \\
\hline
U-Net & 8.6M & 0.660 & \textbf{0.632} & 0.698   \\
\hline
U-Net & 10.9M & \textbf{0.665} & 0.600 & \textbf{0.701}  \\
\hline
Res. U-Net & 2.7M & 0.632 & 0.588 & 0.697  \\
\hline
Res. U-Net & 4.7M & 0.656 & 0.609 & 0.696  \\
\hline
Res. U-Net & 7.4M & 0.557 & 0.541 & 0.645  \\
\hline
R2U-Net & 3.4M & 0.634 & 0.606 & 0.686  \\
\hline
R2U-Net & 6.0M & 0.614 & 0.530 & 0.647  \\
\hline
R2U-Net & 9.4M & 0.631 & 0.572 & 0.666  \\
\hline
Attn. U-Net & 3.1M & 0.517 & 0.439 & 0.627  \\
\hline
Attn. U-Net & 4.5M & 0.529 & 0.465 & 0.586 \\
\hline
\multicolumn{5}{l}{}
\end{tabular}
\label{tab-crossval}
\end{center}
\end{table}

\begin{figure*}
\centering
   {\epsfig{file = 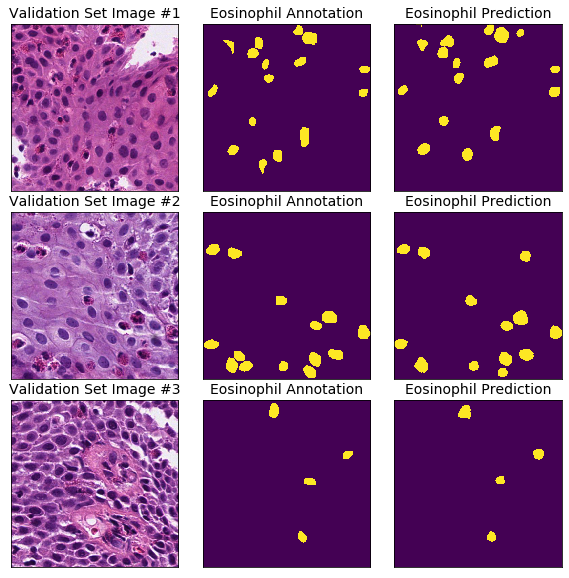, width = 8.5cm}}
  \caption{Examples of eosinophil detections from the U-Net model. On the left, are three input images from the validation set. In the middle, are the ground-truth annotations. On the right, are the prediction masks.}
  \label{fig:example_Eos}
 \end{figure*}

The U-Net model performed better on eosinophil detection than the other more advanced approaches. This could be due to the U-Net being able to accommodate more parameters within GPU memory limits. The 10.9M U-Net had a slightly higher median and maximum test dice set coefficient, but the 8.6M U-Net had a much higher minimum value. We selected the 8.9M U-Net since it was less complex and provided near equivalent accuracy as the 10.9M U-Net. To establish the final 8.9M U-Net model, only training and validation sets were utilized from the 274 total images. The size of the training set was 214 images while the validation set size was 60 images. The resulting validation Dice coefficient after training with a similar process was 0.705. Figure \ref{fig:example_Eos} shows a few examples of eosinophil detections on validation set images.

\subsubsection{Eosinophil Counting and Validation}

Successful eosinophil segmentation is crucial to count cells and generate per-image and per-patient features. With the U-Net segmentation model established, we then developed a post-processing technique to extract the cell counts. First, the output of the U-Net model was converted to a binary segmentation mask using a 0.5 probability threshold. The resultant segmentation masks sometimes contained small contiguous areas or artifacts that were not eosinophils.  Eosinophils are typically 800 contiguous pixels in size, so objects with less than 200 contiguous pixels were removed to reduce false positive detections and still preserve partially-detected eosinophils. The next step was performing an Euclidean distance transformation in an attempt to separate any eosinophils that may overlap \cite{heinz1995linear}. This operation reduced the size of cells from the outside edge so that only the centers were counted. The approach is similar to erosion, but is more computationally efficient. A distance of eight pixels was used as the cutoff, because it reduced the cell size by roughly 75\% and can separate large cell overlaps. Finally, the \textit{measure.label} function was applied from the Skimage package in Python. The \textit{label()} function applied a connected component labeling algorithm to separately label each contiguous object in the prediction mask. The count of unique labels is one of the outputs of this function and thus represents the number of eosinophils detected in that image. 

\begin{figure}
\centering
   {\epsfig{file = 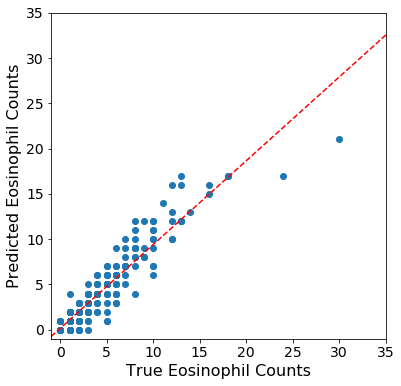, width = 5.5cm}}
  \caption{Eosinophil counting errors. Each point on this scatterplot is one of the 274 EoE patches. The dotted red line represents the linear trend.}
  \label{fig:eo_counts}
 \end{figure}

Since the validation set Dice coefficient was optimized during model training, the entire 274 images dataset was used to evaluate the cell counting capability of segmentation and post-processing. The counting error was estimated by subtracting actual and predicted eosinophil counts. The average eosinophil error over all $512 \times 512$ pixel images was $-0.12$ with a standard deviation of $1.39$ eosinophils. When scaled to the size of an HPF, the average error was roughly $-2$ eosinophils. As shown in Figure \ref{fig:eo_counts}, the negative bias was most pronounced when the true eosinophil count was greater than 20. The requirement to diagnose EoE is 15 or more eosinophils within an area roughly $16\times$ larger, so the bias should not affect diagnosis. When the true eosinophil count was less than 15, minimal over-counting bias was noticed. This was not problematic since false-positive diagnoses and severity estimation are preferred over false-negatives as it avoids under-diagnosing EoE which can lead to worsened patient outcomes \cite{lipka2016impact}. When the number of eosinophils reach over 20 per patch, the most likely cause of under-counting is the inability to separate and count overlapping cells on the prediction mask. 

\subsection{Classification for EoE Diagnosis}

An alternative to the segmentation approach for diagnosis is to extract WSI patches from patients who are EoE-diagnosed or histologically normal and train a CNN classification model to predict EoE versus normal. Training, validation, and test sets were required to develop this model. A total of 36 and 41 patients were used for EoE and normal, respectively. Patches from WSIs for patients were evenly sampled per patient and also sampled so that there was minimal class imbalance in the training and validation sets. Table \ref{tab-classdata} shows the number of patients and patches in each data set.   

\begin{table}[htbp]
\caption{Classification Model Data Sets}
\begin{center}
\begin{tabular}{|c|c|c|c|c|}
\hline
& \multicolumn{2}{|c|}{\textbf{\# of Patients}} & \multicolumn{2}{|c|}{\textbf{\# of Patches}} \\
\cline{2-5}
\textbf{Data Set} & \textbf{EoE}& \textbf{Normal} & \textbf{EoE} & \textbf{Normal} \\
\hline
Training & 21 & 24 & 15,676 & 15,780 \\
\hline
Validation & 8 & 10 & 3,642 & 3,538 \\
\hline
Test & 7 & 7 & 7,631 & 3,474 \\
\hline
\multicolumn{5}{l}{}
\end{tabular}
\label{tab-classdata}
\end{center}
\end{table}

A VGG16 model was trained for 12 epochs with an Adam optimizer and learning rate of $2\times10^{-4}$. The model weights were saved that achieved the best validation accuracy. The highest validation accuracy achieved was 0.994. The model weights with this validation accuracy achieved a test set accuracy of 0.959. As shown in Table \ref{tab-confusion}, the model rarely classified normal patches as EoE, but there were a greater number of false-positive patches. Test set accuracies this high are quite interesting, because it shows that EoE can be accurately diagnosed through just a single $512\times512$ pixel patch. This greatly differs from the typical diagnosis method, so it could possibly be used to discover new features that indicate EoE. In the Results section, we dive deeper into the test set results and discuss the Grad-CAM analysis.

\begin{table}[htbp]
\caption{Test Set Confusion Matrix for Classification Model}
\begin{center}
\begin{tabular}{l|c|c|c|}
\multicolumn{2}{c}{}&\multicolumn{2}{c}{\textbf{Predicted}}\\
\cline{3-4}
\multicolumn{2}{c|}{}& EoE & Normal \\
\cline{2-4}
\multirow{2}{*}{\textbf{True Diagnosis}}& EoE & 7,176 & 455 \\
\cline{2-4}
& Normal & 3 & 3,471 \\
\cline{2-4}
\end{tabular}
\label{tab-confusion}
\end{center}
\end{table}

\section{\uppercase{Results}}

\begin{table*}[ht]
\caption{EoE Patient Categorizations}
\begin{center}
\begin{tabular}{|c|c|c|c|c|c|c|c|}
\multicolumn{2}{c}{}&\multicolumn{6}{c}{\textbf{Treatment Plans}}\\
\cline{3-8}
\multicolumn{1}{c}{} & \textbf{Phenotypes} & \textbf{\textit{PPI-REE}} & \textbf{\textit{Milk Removal}} & \textbf{\textit{4/6 FED}} & \textbf{\textit{Steroids}} & \textbf{\textit{Unk}} & \textbf{\textit{Total}} \\
\hline
\multirow{4}{*}{\textbf{Initial Biopsy}} & \textbf{\textit{PPI}} & 9 & & & & & 9  \\
\cline{2-8}
& \textbf{\textit{Inflammatory}} &  & 1 & 4 & 6 & & 11  \\
\cline{2-8}
& \textbf{\textit{Strictures}} &  &  & 2 & 5 & 1 & 8  \\
\cline{2-8}
& \textbf{\textit{Total}} & 9 & 1 & 6 & 11 & 1 & 28  \\
\hline\hline
\multirow{4}{*}{\textbf{Not Initial}} & \textbf{\textit{PPI}} &  & & & & & 0  \\
\cline{2-8}
& \textbf{\textit{Inflammatory}} &  & 2 & 3 & 8 & 4 & 17  \\
\cline{2-8}
& \textbf{\textit{Strictures}} &  &  &  & 1 & 2 & 3  \\
\cline{2-8}
& \textbf{\textit{Total}} & 0 & 2 & 3 & 9 & 6 & 20  \\
\hline
\end{tabular}
\label{tab-EoEpatients}
\end{center}
\end{table*}

In this section, we present findings generated from the image segmentation and classification approaches. For image segmentation, the model was applied to a larger set of EoE patients with completed retrospective chart reviews. This led to patient-level features for characterization of EoE that were analyzed to find linkages with treatment and clinical phenotypes. The image classification model produced an accurate fit on small WSI patches. We looked further into a single patient who had a large number of misclassified patches and assessed Grad-CAMs to understand the differences within their biopsy tissue sample.

\subsection{Image Segmentation Results}

A total of 44 EoE patients with completed retrospective chart reviews and 57 histologically normal patients were used in this section to assess diagnostic capability and relationships with other biopsy characteristics. The EoE patients were categorized into two major phenotypes: clinical and treatment-based. The clinical phenotypes were based on patients developing strictures vs. those with the disease remaining inflammatory vs. PPI-REE. Treatment phenotypes included patients with disease that were responsive to dairy-only (milk) elimination vs. 4 to 6 food elimination diet (4/6 FED) vs. steroids vs. unknown (where treatment response was unclear). Biopsies from these patients were all retrieved at the same university hospital at the time of initial EoE diagnosis. To assess the linkages between initial biopsy and phenotypes or treatments it was important to isolate the patients that have true initial biopsies. Table \ref{tab-EoEpatients} shows the how many patients were distributed in each bin for all possible categories. 

Various statistics were calculated for each patient to represent the severity and extent of EoE. Patients had up to three WSIs each depending on how many different esophageal locations were sampled during the endoscopy. For all results, the HPFs from all WSIs were aggregated for each patient and not separated by esophagus location. HPFs were not ideal for collecting information about localized concentrations of eosinophils. Therefore, statistics were also collected for $512 \times 512$ pixel patches to determine if there were any micro-level trends that differ from the macro-level. Due to the small sample sizes within the treatment and clinical phenotype categories after splitting for initial biopsy, the non-parametric Wilcoxon rank sum test was used to test all statistical differences between each category pair \cite{wilcoxon1970critical}. The statistics are also highly correlated with each other within this small set of patients, so no multivariate testing was performed. Multivariate testing could be possible when more EoE patient response data becomes available. The following list details some of the eosinophil statistics that were captured:

\begin{itemize}
    \item Maximum eosinophils in HPFs
    \item Average eosinophil count over all HPFs
    \item Average eosinophil size (pixels)
    \item Percent of HPFs with zero eosinophils
    \item Percent of HPFs with $\leq 5$ eosinophils
    \item Percent of HPFs with $\geq$ 15, 30, 60  eosinophils
    \item Maximum eosinophils in $512\times512$ patches
    \item Average eosinophil count over all patches
    \item Percent of patches with zero eosinophils
    \item Percent of patches with $\geq$ 5, 10, 15 eosinophils
\end{itemize}

\subsubsection{Diagnosis, Severity, and Extent}

Using the criteria of 15 or more eosinophils within an HPF, patients were able to be diagnosed via results obtained from the image segmentation model. Table \ref{tab-diagnosis} shows the classification results on the 44 and 57 patients for EoE and normal, respectively. The overall accuracy of this approach was $99.0\%$. The sensitivity was $100\%$ and the specificity was $98.2\%$. This demonstrates that the  automated approach was able to adequately diagnose EoE and thus could be a useful tool to assist pathologists. 

\begin{table}[htbp]
\caption{Confusion Matrix for EoE Diagnosis}
\begin{center}
\begin{tabular}{l|c|c|c|}
\multicolumn{2}{c}{}&\multicolumn{2}{c}{\textbf{Predicted}}\\
\cline{3-4}
\multicolumn{2}{c|}{}& EoE & Normal \\
\cline{2-4}
\multirow{2}{*}{\textbf{True Diagnosis}}& EoE & 44 & 0 \\
\cline{2-4}
& Normal & 1 & 56 \\
\cline{2-4}
\end{tabular}
\label{tab-diagnosis}
\end{center}
\end{table}

Additionally, EoE severity can be predicted using the eosinophil statistics. Due to the difficulty of manually counting all eosinophils, pathologists typically do not continue counting eosinophils after reaching about 60 counted in an HPF \cite{collins2017cincyeoe}. This was not a limitation for the automated deep learning approach. The maximum eosinophil count can be counted as an unbounded continuous variable which provides more fidelity over the manual method. 

\begin{figure}
\centering
   {\epsfig{file = 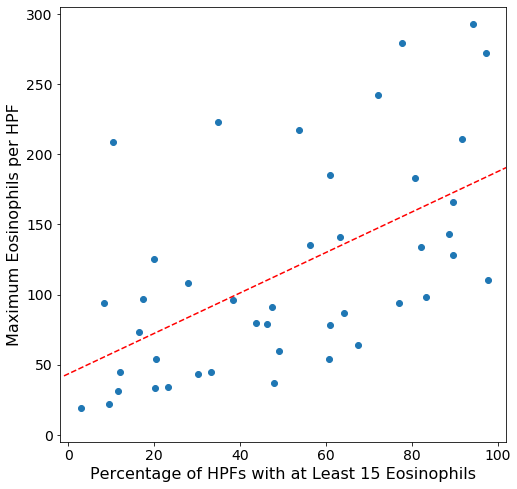, width = 5.5cm}}
  \caption{EoE severity versus extent. The $x$-axis represents extent, while the $y$-axis represent severity. The dotted red line is the linear trend between the two factors.}
  \label{fig:severity_extent}
 \end{figure}

Similar difficulties were present when manually estimating EoE extent. The image segmentation approach obtained counts from all HPF samples and estimated the extent by calculating the percentage of HPFs that exceeded a certain criteria. Figure \ref{fig:severity_extent} shows a scatterplot of each EoE patient's maximum eosinophil per HPF and percentage of HPFs with 15 or more eosinophils. Severity and extent appeared to correlate, but there were patients who deviated from the trend-line. 

\subsubsection{Linkages with Treatment Phenotypes}

The goal of this subsection was to identify linkages between collected eosinophil statistics at the time of initial biopsy and optimal treatment plans. Treatments that the patients responded to in our dataset were assessed at 6 or more months of follow up via retrospective chart review. As shown in Table \ref{tab-EoEpatients}, the sample sizes for initial biopsy patients were small but there was enough data points within 4/6 FED, PPI-REE, and steroids to generate statistical significance.

\begin{figure}
\centering
   {\epsfig{file = 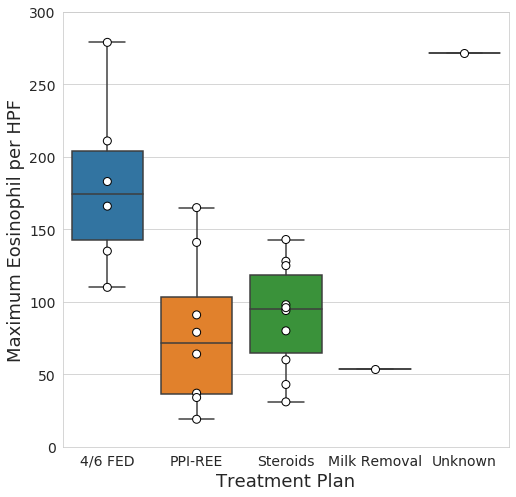, width = 6cm}}
  \caption{Maximum eosinophils over HPFs per patient by treatment plan. Patients with high EoE severity tend benefit most from 4 to 6 food elimination treatment.}
  \label{fig:max_eo_treatment}
 \end{figure}

Figure \ref{fig:max_eo_treatment} shows the maximum eosinophil count for each patient based on their optimal treatment plan. The maximum count was generated from all HPFs within each patient's WSIs. The milk removal category did not have enough samples to evaluate and there was also one unknown treatment. The 4/6 FED treatment appeared to correlate with higher EoE severity. The statistical test revealed that 4/6 FED had a significantly higher maximum eosinophil count than PPI-REE and steroids with p-values of 0.012 and 0.006, respectively.   

\begin{figure}
\centering
   {\epsfig{file = 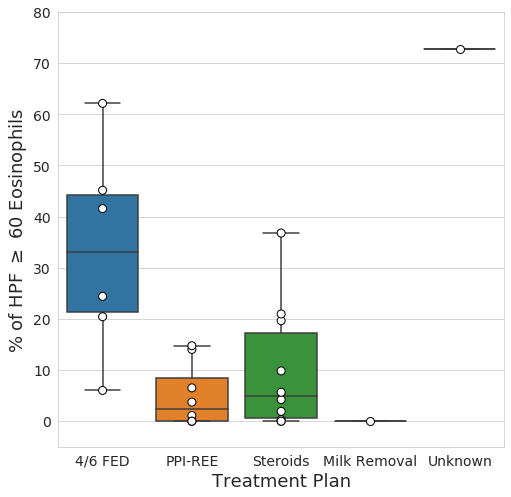, width = 6cm}}
  \caption{Percentage of HPFs with $\geq$ 60 eosinophils. Patients with high EoE extent tend to benefit most from 4 to 6 food elimination treatment.}
  \label{fig:hpf_60_treatment}
\end{figure}

Figure \ref{fig:hpf_60_treatment} shows the percentage of HPFs that contain at least 60 eosinophils by treatment plan. The 4/6 FED treatment also appeared to correlate with higher EoE extent. There were similar trends with lower eosinophil thresholds such as 15 or 30, but 60 showed a clearer difference between the treatments. The statistical test revealed that 4/6 FED had a significantly higher percentage of HPFs with $\geq$ 60 eosinophils than PPI-REE and steroids with p-values of 0.008 and 0.015, respectively.

\begin{figure}
\centering
   {\epsfig{file = 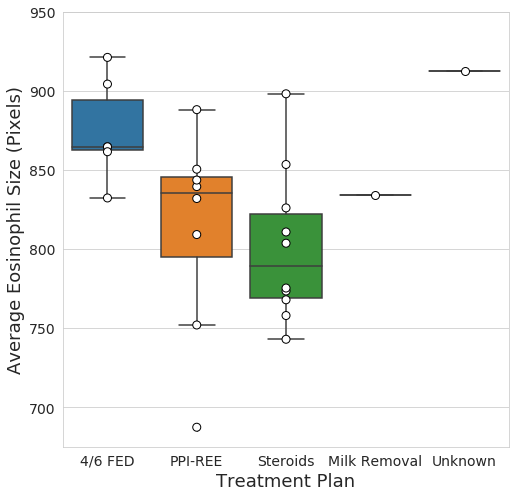, width = 6cm}}
  \caption{Average eosinophil size in pixels by treatment plan. Patients that have larger eosinophils tend to benefit most from 4 to 6 food elimination treatment.}
  \label{fig:eo_size_treatment}
 \end{figure}

Figure \ref{fig:eo_size_treatment} shows the average eosinophil size (pixels) by treatment plan. The average eosinophil size is calculated by summing the number of eosinophil-detected pixels in each HPF and then dividing by the predicted count. Not only does 4/6 FED treatment seem to link with patients with higher EoE severity and extent, but also with the actual size of the eosinophils. The statistical test revealed that 4/6 FED had a significantly higher average eosinophil size than PPI-REE and steroids with p-values of 0.033 and 0.008, respectively. 

There were other variables such as the average eosinophil count and similar statistics at the 512 $\times$ 512 pixel patch that also showed 4/6 FED as significantly different from PPI-REE or steroids treatment.

\subsubsection{Linkages with Clinical Phenotypes}

\begin{figure}
\centering
   {\epsfig{file = 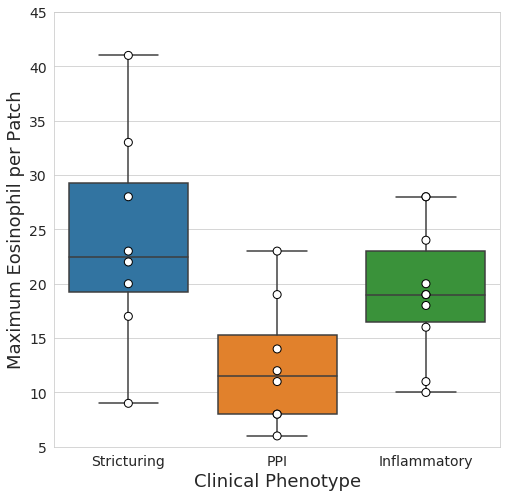, width = 6cm}}
  \caption{Maximum eosinophil count over patches per patient by phenotype. Patients with a low eosinophil counts in smaller regions tend to be linked with the PPI phenotype.}
  \label{fig:max_patch_phenotype}
\end{figure}

The goal of this subsection was similar to the previous as we searched for linkages between initial biopsy features and a patient's clinical phenotype. While the sample sizes were still low, the categories were fairly well-balanced with all three ranging between 8 and 11. Figure \ref{fig:max_patch_phenotype} shows the maximum eosinophil count per patient using 512 $\times$ 512 patches and by phenotype. The treatment phenotype findings were similar for both HPFs and patches, but the clinical phenotype only generated significant differences at the ``micro-level''. Therefore, the indicators of phenotypes at the time of initial biopsy and diagnosis may only exist when examining eosinophils at a highly localized level and not through HPFs. The statistical test revealed that PPI-REE had a significantly lower maximum eosinophil count per patch than strictures and inflammatory with p-values of 0.021 and 0.049, respectively.  

Figure \ref{fig:max_patch_phenotype} shows the maximum eosinophil count per patient using 512 $\times$ 512 patches by clinical phenotype. This shows another example of how patients with the PPI-REE phenotype rarely ever exceeded ten eosinophils within a 512 $\times$ 512 patch. The median percentage for PPI was less than 1\%, while the other two had medians at roughly 4\%. The statistical test revealed that PPI-REE had a significantly lower percentage of patches with $\geq$ 10 eosinophils than strictures and inflammatory with p-values of 0.034 and 0.011, respectively.  
 
\begin{figure}
\centering
   {\epsfig{file = 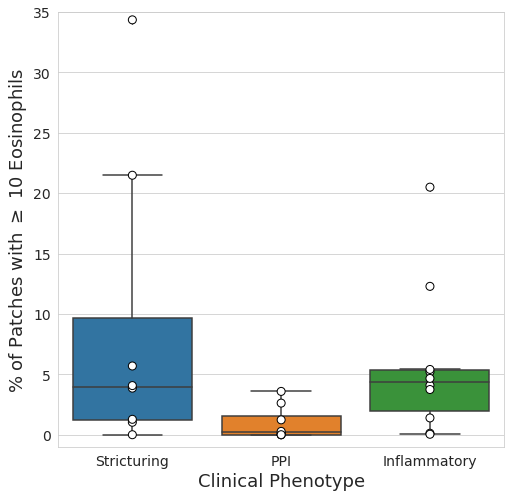, width = 6cm}}
  \caption{Percentage of patches with $\geq$ 10 eosinophils per patient by phenotype. Patients with very low percentages of patches with a large number of eosinophils tend to have the PPI-REE phenotype}
  \label{fig:patch_10_phenotype}
\end{figure}

\subsection{Image Classification Results}

\begin{figure*}
\centering
   {\epsfig{file = 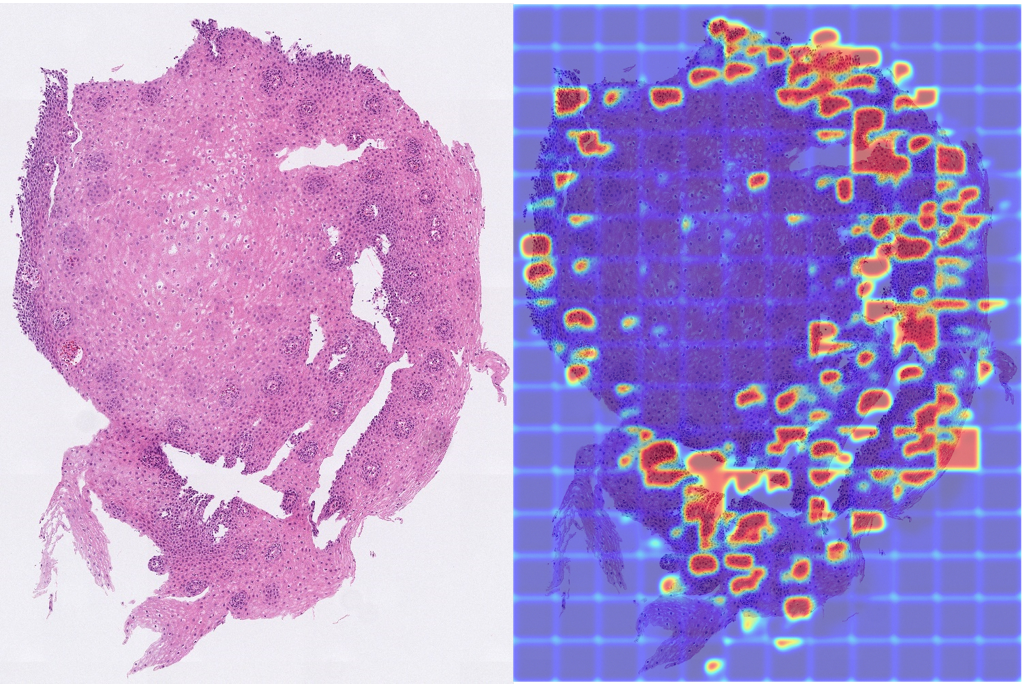, width = 10.5cm}}
  \caption{Grad-CAMs are applied to patches over a section of tissue from patient E-136's WSI. On the left, is the section of WSI tissue. On the right is the same crop of tissue, but with the Grad-CAM overlaid. This was the only EoE patient in the test set that had large areas of tissue predicted as normal.}
  \label{fig:gradcam}
 \end{figure*}

As discussed in the Methodology section, a VGG16 CNN model was developed with training and validation data sets and then evaluated on a test set. The training and validation sets contained EoE patients from the 4/6 FED, PPI-REE, and steroids treatment plans, while the test set EoE patients were from the milk removal and unknown treatments. Overall, the prediction accuracies were high across all three data sets, but there was one interesting anomaly in the test set results. Table \ref{tab-class_preds} shows the classification results from the VGG16 model by each patient from the test set. All patients had almost all of their patches correctly classified except for EoE patient E-136. 

\begin{table}[htbp]
\caption{Classification Predictions by Patient}
\begin{center}
\begin{tabular}{|c|c|c|c|}
\cline{3-4}
\multicolumn{2}{c}{} & \multicolumn{2}{|c|}{\textbf{Prediction}} \\
\hline
\textbf{Truth} & \textbf{Patient}& \textbf{\textit{EoE}} & \textbf{\textit{Normal}} \\
\hline
\multirow{7}{*}{\textbf{\textit{EoE}}} & E-29 & 187 & 1 \\
\cline{2-4}
& E-103 & 701 &   \\
\cline{2-4}
& E-116 & 798 &  \\
\cline{2-4}
& E-123 & 1,567 & 17 \\
\cline{2-4}
& E-124 & 784 & 2 \\
\cline{2-4}
& E-136 & 808 & \textcolor{red}{\textbf{435}} \\
\cline{2-4}
& E-201 & 2,331 &  \\
\hline

\multirow{7}{*}{\textbf{\textit{Normal}}} & N16-38 &  & 450 \\
\cline{2-4}
& N16-39 & 3 & 1,317  \\
\cline{2-4}
& N16-40 &  & 355 \\
\cline{2-4}
& N16-41 &  & 651 \\
\cline{2-4}
& N16-42 &  & 285  \\
\cline{2-4}
& N16-43 &  & 381 \\
\cline{2-4}
& N16-44 &  & 32  \\
\hline

\end{tabular}
\label{tab-class_preds}
\end{center}
\end{table}

About one-third of the WSI patches for E-136 were misclassified as normal, while other patients had almost 100\% correctly classified EoE patches. The fact that predictions varied within one patient's WSI was one piece of supporting evidence that the model was not utilizing extraneous features to classify EoE vs. normal. Examining patient E-136's tissue sample may be the key to determining which biopsy features were utilized by the deep learning model. Grad-CAMs can be utilized to visualize the important features used in a CNN model. Grad-CAMs were produced at the patch-level to cover E-136's entire WSI in order to assess which parts of the tissue were considered EoE or normal. Pathologists examined the Grad-CAMs and determined whether there was a consistent trend associated with the tissue and a prediction class. The WSI-level Grad-CAMs were highlighting more areas for images that had a higher number of eosinophils. Further, the heatmaps were also focusing on areas with cellular crowding and bottom most (basal) layer of epithelium with images having dilated intercellular spaces. These increased (dilated) intercellular spaces indicate underlying cellular edema taking place due to the inflammation caused by the disease \cite{collins2017cincyeoe}. 

Figure \ref{fig:gradcam} shows an example of a large tissue crop from one on the WSIs from patient E-136. There were large sections in the middle of the tissue sample that the model considers normal (blue), while other areas are predicted EoE (red). After analyzing patient E-136 and comparing to other patient's Grad-CAMs, it was determined the model was classifying patches EoE or normal likely because of features other than eosinophils such as cellular crowding and basal layer of the epithelium. Another important aspect was that the classification model was not utilizing eosinophils to diagnose EoE. The section of tissue in Figure \ref{fig:gradcam} was very large (8000 $\times$ 6000 pixels) and about half was predicted as EoE, but there were only three eosinophils detected within this area. This amount of eosinophils is far short of the typical diagnostic criteria.

\section{\uppercase{Conclusion}}
\label{sec:conclusion}

In this paper, we present a novel approach for diagnosing EoE, understanding more about biopsy tissue level features, and linking EoE biopsy features with treatment and clinical phenotypes. This was the first time deep learning computer vision was applied to diagnose EoE and detect eosinophils in biopsy tissue samples. The eosinophil segmentation is medically critical since it can aide pathologists in diagnosis by automatically providing peak eosinophil counts and knowledge of other areas requiring direct focus and attention. The automated assessment of eosinophil count can also be associated with disease severity, progression, and medically-relevant treatment and clinical phenotypes. The trained U-Net image segmentation model detected eosinophils within an adequate accuracy. 

Most of the patient-level eosinophil statistics cannot be realistically captured without an automated approach. These statistics can be used to explain the EoE severity and extent or to predict clinical and treatment phenotypes. Statistical analysis revealed that patients who eventually responded to the 4/6 FED treatment plan had high severity and extent of EoE at the time of initial biopsy. The PPI-REE phenotype was found to relate with lower severity (decreased eosinophil counts) and extent at the smaller patch level. 

In conjunction with image segmentation, image classification was used to learn more about how biopsy features relate to EoE. A VGG16 CNN model was trained on a data set of EoE and normal WSI patches and achieved highly accurate results. The model's performance was interesting since it was able to predict EoE using only small patches and did not depend solely on eosinophils for diagnosis. One EoE patient in particular had a large section of tissue predicted as normal and this was because it had less areas of eosinophils which lead to less activations mapped by Grad-CAMs but also features other than eosinophils such as cellular crowding and basal epithelial layer with some images have dilated intercellular spaces were being highlighted. Basal layer epithelium identification and its thickness along with presence of dilated intercellular spaces representing cellular edema are signs of inflammation caused by EoE and have been proposed as possible diagnostic features of the disease \cite{collins2017cincyeoe}.  

There are many avenues for future research in this area. Increasing eosinophil annotations will further improve prediction performance and enable the model to be more robust to even slight variations in new biopsy imagery. The sample size of patient-level data was small due to which most of the statistical analysis was limited. There is currently a plan in place to greatly increase the sample size of patients with completed chart reviews. With additional data, we will also be able to analyze eosinophil trends by the esophageal tissue sample location. The Grad-CAM review can be considered subjective, thus integrating Grad-CAMs with another approach that can quantify cellular-level features will reduce human-injected bias. We are in the process of developing an unsupervised segmentation approach that will also be reviewed by pathologists to automate patch-level feature generation.   

\bibliographystyle{apalike}
{\small
\bibliography{main}}

\end{document}